\documentclass[preprint2]{emulateapj}  

\IfFileExists{srcltx.sty}{\usepackage[active]{srcltx}}  

\shorttitle{New Limits on Sterile Neutrinos}
\shortauthors{Loewenstein et al.}

\begin{document}

\title{New Limits on Sterile Neutrinos from {\em Suzaku} Observations of
the Ursa Minor Dwarf Spheroidal Galaxy} 
\author{Michael Loewenstein\altaffilmark{1,2}, Alexander
Kusenko\altaffilmark{3,4}, Peter L. Biermann\altaffilmark{5,6,7}}
\email{Michael.Loewenstein-1@nasa.gov}
\altaffiltext{1}{Department of Astronomy, University of Maryland,
College Park, MD.}
\altaffiltext{2}{CRESST and X-ray Astrophysics Laboratory NASA/GSFC,
Greenbelt, MD.}
\altaffiltext{3}{Department of Physics and Astronomy, University of
California, Los Angeles, CA 90095-1547, USA}
\altaffiltext{4}{Institute for the Physics and Mathematics of the Universe,
University of Tokyo, Kashiwa, Chiba 277-8568, Japan}
\altaffiltext{5}{Max-Planck-Institut für Radioastronomie, Bonn,
Germany.}
\altaffiltext{6}{Department of Physics and Astronomy, University of
Bonn, Bonn, Germany.}
\altaffiltext{7}{Department of Physics and Astronomy, University of
Alabama, Tuscaloosa, AL}

\received{2008 December 14}\accepted{2009 May 19}

\slugcomment{To be published in the {\it Astrophysical Journal}
Vol. 698, UCLA/08/TEP/32}


\begin{abstract}
We present results of our search for X-ray line emission associated
with the radiative decay of the sterile neutrino, a well-motivated
dark matter candidate, in {\it Suzaku} Observatory spectra of the Ursa
Minor dwarf spheroidal galaxy. These data represent the first deep
observation of one of these extreme mass-to-light systems and the
first dedicated dark matter search using an X-ray telescope. No such
emission line is positively detected, and we place new constraints on
the combination of the sterile neutrino mass, $m_{\rm st}$, and the
active-sterile neutrino oscillation mixing angle, $\theta$. Line flux
upper limits are derived using a maximum-likelihood-based approach
that, along with the lack of intrinsic X-ray emission, enables us to
minimize systematics and account for those that remain. The limits we
derive match or approach the best previous results over the entire
1--20 keV mass range from a single {\it Suzaku} observation. These are
used to place constraints on the existence of sterile neutrinos with
given parameters in the general case and in the case where they are
assumed to constitute all of the dark matter. The range allowed
implies that sterile neutrinos remain a viable candidate to make up
some -- or all -- of the dark matter and also explain pulsar kicks and
various other astrophysical phenomena.
\end{abstract}



\section{Introduction}

The observational evidence that nonbaryonic dark matter comprises most
of the mass in the universe is extremely strong, but the nature of
dark-matter particles remains a mystery.  No particle included in the
Standard Model of particle physics has the characteristics required to
explain the dark matter.  In its original formulation, the Standard
Model described all known particles, including neutrinos, which were
assumed to be massless.  The discovery that neutrinos have mass has
forced one to go beyond this minimal model: a modest extension of the
Standard Model by gauge singlet fermions can accommodate the neutrino
masses.  One or more of these gauge-singlet fermions can have Majorana
masses below the electroweak scale, in which case they appear as
sterile neutrinos in the low-energy theory.  If one of the sterile
neutrinos has mass in the 1--20~keV range and has small mixing angles
with the active neutrinos (as expected), such a particle is a
plausible candidate for dark matter~\citep{dw94}.  The same particle
could be produced in a supernova explosion, and its emission from a
cooling neutron star could explain the pulsar kicks
\citep{ks97,Fuller03,k04,kmm}, could facilitate core collapse
supernova explosions~\citep{Fryer,hf07} and can affect the formation
of the first stars~\citep{bk06,s07} and black holes~\citep{mb05,mb06}.
Therefore, there is a strong motivation to search for signatures of
sterile neutrinos in this mass range.

The most promising way to discover (or rule out) relic sterile
neutrinos is with the use of X-ray telescopes (XRTs), such as {\em Suzaku}.  
The sterile neutrinos can decay radiatively~\citep{pw,barger} and
produce a lighter neutrino and a photon amenable to X-ray
observation~\citep{afp,aft,dh}.  Since this is a two-body decay, a
line is expected in the X-ray spectrum.  The dwarf spheroidal galaxies
are ideal targets for these observations because of their proximity,
high dark matter density \citep{s08a}, and absence of additional
obscuring X-ray sources (see below). The Ursa Minor and Draco dwarf
spheroidals are the optimal targets in this class for large
field-of-view X-ray spectroscopic investigation, based on their
average dark matter surface densities. From {\it XMM-Newton}
observations of the Ursa Minor dwarf spheroidal, sterile neutrino line
flux limits comparable to the best previous constraints were derived
-- even though strong background flaring limited the amount of good
observing time to 7 ks \citep{bnr}.  The {\em Suzaku} Observatory
\citep{m07} provides the most sensitive instruments for current
searches for weak sterile neutrino radiative decay lines in the $\sim
0.5-10$ keV bandpass because of its low and stable background
\citep{ta08}, and the relatively sharp spectral resolution of its CCD
spectrometers \citep{k07,n08}.  In this paper, we discuss our analysis
of {\it Suzaku} observations of the Ursa Minor system, considerably
improving on the {\it XMM-Newton} constraints; a companion article on
the Draco system is in preparation. These represent the first
substantial X-ray data of such extremely dark-matter-dominated
systems.

\subsection{Context}

Sterile neutrinos may be produced through non-resonant oscillations,
as proposed by Dodelson and Widrow \citep{dw94}, and via various other
mechanisms~\citep{d02,sf99,st,k06,kadota,pk08}.  A generic prediction
is that relic sterile neutrinos must decay into a lighter neutrino and
a photon. Since this is a two-body decay of a particle with mass
$m_{\rm st}$, decay photons produce an intrinsically narrow line with
energy $E_\gamma = m_{\rm st}/2$ that can be detected using X-ray
spectroscopy \citep{aft}. While the decay timescale, $\Gamma_{\rm
st}^{-1}$, greatly exceeds the Hubble time, the emissivity may reach
detectable levels in regions of large dark matter concentration. Upper
limits on $\Gamma_{\rm st}$ from X-ray observations of the cosmic
background \citep{ak06,b06a,a07}, galaxy clusters
\citep{b06b,r07,brm}, and M31 \citep{w06,r07,birs} may be complemented
and improved by observing dwarf spheroidal galaxies \citep{b06c}.

There is no consensus as to whether dark matter is cold or warm based
on observations of small-scale structure. Cold and warm dark matter
reproduce large-scale structure equally well, but differ in their
predictions on scales comparable to the free-streaming length.
Ly$\alpha$ forest clouds provide a tracer population of the
perturbation spectrum on the appropriate (non-linear) scale that may
be used to place an upper limit on the free-streaming length
\citep{n00}, and have been interpreted as implying that the mass of
the dark matter sterile neutrino, $m_{\rm st}$, must be greater than
$\sim 8~$keV \citep{v08,blrv} for production exclusively by the
Dodelson-Widrow (DW) non-resonant mixing mechanism. However, $N$-body
simulations and observations of dwarf spheroidal galaxies imply a
lower mass \citep{g07,s07a}. In scenarios with alternative production
mechanisms that do not involve oscillations, sterile neutrinos may
effectively be colder, allowing masses as low as 3~keV to be
consistent with the Ly$\alpha$ observations~\citep{k06,pk08,pet08}.
In this case, the photometric and kinematic data from dwarf spheroidal
galaxies~\citep{g07,s07a} can be satisfied by dark matter in the form
of sterile neutrinos with mass between 0.5~keV and
1.3~keV~\citep{b08}.  Constraints on sterile neutrino parameters may
be further relaxed in mixed dark matter models \citep{p07,blrv} and in
models where sterile neutrinos are produced in resonant oscillations
in the presence of significant lepton asymmetries
\citep{sf99,ls08}. Since the sterile neutrino distribution function
may substantially deviate from thermal equilibrium
\citep{a06a,a06b,pet08,b08}, simple scaling arguments and alterations
to the power spectrum (e.g., imposing a cutoff) do not suffice. New
large-scale structure formation simulations with the appropriate
revisions are required to re-evaluate the constraints from Ly$\alpha$
data \citep{ls08,b08}.  Moreover, their interpretation relies on a
thorough understanding of the uncertain thermal and ionization
evolution of the intergalactic medium, which may be affected by the
properties and interactions of dark-matter
particles~\citep{bk06,gao,reviewLya}. Hence, we present limits solely
derived from our X-ray data.

The sterile neutrino mass and the mixing angle $\theta$ determine the
rate of radiative decay to active neutrinos and X-ray photons:
$\nu_s\rightarrow\gamma \nu_a$.  The radiative decay width is equal
to~\citep{pw}
\begin{eqnarray}
 \Gamma_{\nu_s\rightarrow\gamma \nu_a} &=& \frac{9}{256\pi^4}\,
 \alpha_{_{\rm EM}}\, {\rm G}_{_{\rm F}}^2 \, \sin^2 \theta \, m_{\rm
 s}^5 \nonumber \\ & = & \frac{1}{1.8\times 10^{31}\, {\rm s}}\ \left(
 \frac{\sin^2 \theta}{10^{-10}} \right) \ \left( \frac{m_{\rm s}}{\rm
 keV}\right)^5,
\end{eqnarray}
where $ \alpha_{_{\rm EM}}=1/137.0$ and $G_{_{\rm F}}=1.2\times
  10^{-5} {\rm GeV}^{-2}$.

For a given production mechanism, one may be able to eliminate one of
these two parameters by requiring that the total density match that of
dark matter in the universe as inferred from observations.  However,
given the various possible mechanisms for producing sterile neutrinos,
and the possibility that sterile neutrinos may account for some -- but
not all -- dark matter \citep{p07}, one should keep an open mind and
search all the available parameter space in the $m_{\rm st}-\theta$
plane.

\section{Data Analysis}

\subsection{Observations and Initial Reduction}

Ursa Minor (OBSID = 802052010) was observed with {\it Suzaku}
\citep{m07} between 2007 April 5 18:33:24 and 2007 April 6 8:08:14
(UT) with an on-source exposure time of 70.90 ks, 57.5 (13.4) ks in a
$5\times 5$ ($3\times 3$) editing mode (this refers to the pixel area
utilized in event identification that is included in the telemetry;
see Koyama et al. 2007). At the time of observation, three co-aligned,
$17'.8\times 17'.8$ field-of-view X-ray Imaging Spectrometer (XIS) CCD
cameras \citep{k07} -- two front-illuminated (FI: XIS0 and XIS3) and
one back-illuminated (BI:XIS1) -- were operational. The BI chip is
more sensitive below 1 keV but has a much higher internal background,
especially above 7 keV, so that BI and FI CCDs complement each other
for broad-band studies such as ours. Each XIS lies in the focal plane
of an XRT with a $2'$ half-power diameter \citep{se07}.

{\it Suzaku} guest observers are provided with unfiltered and
pre-screened event files. We initiate our data reduction with the Ursa
Minor unfiltered event files. These underwent Version 2.0.6.13
pipeline processing that, for each event, assigns a pixel quality
status, calculates the sky coordinates using the current aspect
solution and correcting for spacecraft wobble \citep{u08a}, and
converts pulse height amplitude (PHA) to PHA Invariant (PI) values
that map to photon energy in a one-to-one manner. Observations were
conducted utilizing the space-row charge injection (SCI) technique
that reverses the degradation in energy resolution caused by
accumulated radiation damage and affects the conversion from the PHA
channel to energy \citep{n08}. Version 2 pre-processing enables one to
properly account for the effect of SCI on instrument characteristics
and performance \citep{u08b}. We reprocess the unfiltered event files
by hand in order to apply updated calibration data and software.  Our
analysis generally follows the procedures outlined in ``The Suzaku
Data Reduction
Guide,''\footnote{http://heasarc.gsfc.nasa.gov/docs/suzaku/analysis/abc/}
as implemented in HEAsoft version
6.5.1\footnote{http://heasarc.gsfc.nasa.gov/docs/software/lheasoft/}. We
recalculate PI values and grades, select event grades (0, 2, 3, 4, 6)
that correspond to X-ray photon events, filter on pixel status
(eliminating bad charge transfer efficiency columns, and rows
invalidated by the charge injection process) and select good time
intervals (GTI) based on pointing, data and telemetry rates, SAA
proximity (``$\rm{SAA}\_ {\rm HXD}\equiv 0, \rm{T}\_ \rm{SAA}\_
\rm{HXD}> 436$''), and proximity to the Earth's limb and illuminated
Earth (``$\rm{ELV}> 5, \rm{DYE}\_ \rm{ELV}> 20$''). In addition,
telemetry-saturated frames and calibration source photons are screened
out and hot and flickering pixels are removed. Finally, we accept only
GTI where the revised geomagnetic cut-off rigidity COR2$>4$, thus
eliminating intervals with the highest particle background level
\citep{ta08} without compromising overall statistical accuracy
(experiments with more stringent criteria -- COR2$> 6$, COR2$> 8$ --
yielded consistent results, but with significantly poorer
statistics). $5\times 5$ event files are converted to a $3\times 3$
mode format, and merged with the $5\times 5$ event files.

\subsection{Spectral Analysis Procedure}

\subsubsection{Generation of Spectra and Spectral Response}

The extraction region used to create source and particle background
spectra is constructed to include the entire detector fields-of-view,
excluding an $8'$ diameter circular region centered on the one bright
point source in the field. This radius was found to balance the
desires to avoid contamination of the diffuse emission from the wings
of the point-spread function and to maximize the statistical accuracy
through inclusion of regions dominated by diffuse emission.

The spectral redistribution matrices ({\bf rmf}) are generated based
on the characteristics and configuration of the instruments at the
time of observation. The {\bf rmf} and spectral files are binned to
2048 channels. The effective area functions ({\bf arf}) are generated
via a Monte Carlo ray-tracing program that calculates the efficiency
of detecting a photon originating from a specified source geometry and
surface brightness distribution within a specified spectral extraction
region \citep{i07}. We input 2,000,000 simulation photons per energy
bin from a uniform source of radius $20'$, cast the spectral
extraction regions into detector coordinates, and adopt a single {\bf arf} 
file for all components (the Ursa Minor dark matter density is
sufficiently flat over this radius that any correction to such a
component is small). The resulting {\bf arf} file is scaled such that
the flux calculated for spectral models corresponds to the full $20'$
circular region.

The spectra from the FI chips, XIS0 and XIS3, are co-added and a
weighted XIS0+3 response function calculated.

\subsubsection{Background Components}

Since dwarf spheroidal galaxies are intrinsically weak X-ray emitters
(see below), their X-ray spectra are dominated by internal and
astrophysical background components. The XIS background includes
contributions from non-X-ray charged particle background (NXB),
galactic X-ray background (GXB), and (extragalactic) cosmic X-ray
background (CXB).

The NXB component is estimated from observations of the night earth
taken in the SCI mode within 150 days of the starting or ending dates
of the Ursa Minor observation \citep{ta08}. The NXB event list in that
time interval undergoes the identical screening as the source data, is
sorted by geomagnetic cut-off rigidity, and weighted according to the
cut-off rigidity distribution in the source event file \citep{ta08}.
The estimated NXB spectra includes only those events collected in the
region on the detector from which the source spectrum is
extracted. XIS0 and XIS3 NXB spectra are co-added. The effective NXB
spectra exposure times and count rates for the XIS0+3 (XIS1) are 1.1
(0.55) million seconds and 0.062 (0.090) counts per second,
respectively, where the count rates are integrated over the 0.45-10.4
(0.45-7.6) keV bandpass (see below).

The GXB may be characterized by a two-temperature thermal plasma
(corresponding to halo and local hot bubble contributions) and the CXB
by a single power law. We find that this model produces a good global
simultaneous fit ($\chi^2=1205$ for 1232 degrees of freedom) to the
NXB-subtracted, binned, 0.45-10.4 keV XIS0+3 and 0.45-7.6 keV XIS1
spectra (Figures 1 and 2), where the two spectra are fitted by the
identical model save for a constant multiplicative factor. Additional
components are not required; however, allowing non-solar oxygen
abundance does improve the fit. We find that the resulting parameters
are consistent with the spectrum extracted from the {\it ROSAT}
All-Sky
Survey\footnote{http://heasarc.gsfc.nasa.gov/cgi-bin/Tools/xraybg/xraybg.pl}.

\begin{figure}
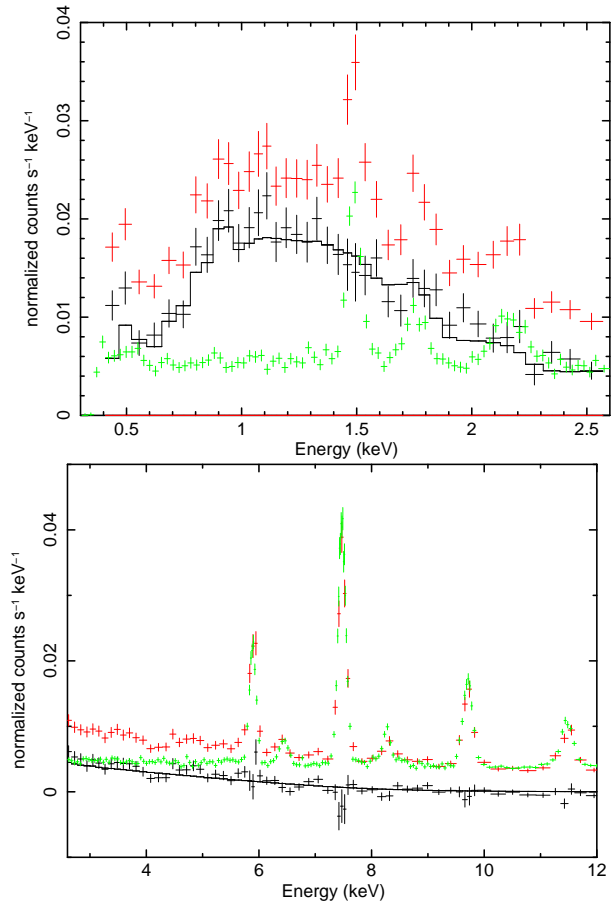

\centering
\includegraphics[scale=0.33,angle=-90]{fig1a.eps}
\hfil 
\centering 
\includegraphics[scale=0.33,angle=-90]{fig1b.eps}
\hfil
\centering
\caption{Low-energy ($<2.6$ keV; {\bf above}) and high energy ($>2.6$
keV; {\bf below}) XIS0+3 spectra. Shown are total (red symbols), NXB
(green symbols), and total$-$NXB (black symbols). The best-fit CXB+GXB
model to the NXB-subtracted spectra is shown by the black histogram.
The best-fit was obtained by simultaneously fitting the 0.45-10.4 keV
XIS0+3 and 0.45-7.6 keV XIS1 spectra; the extension of the model, as
well as that of the data, is plotted over the full 0.3-12 keV bandpass.}
\end{figure}

\begin{figure}
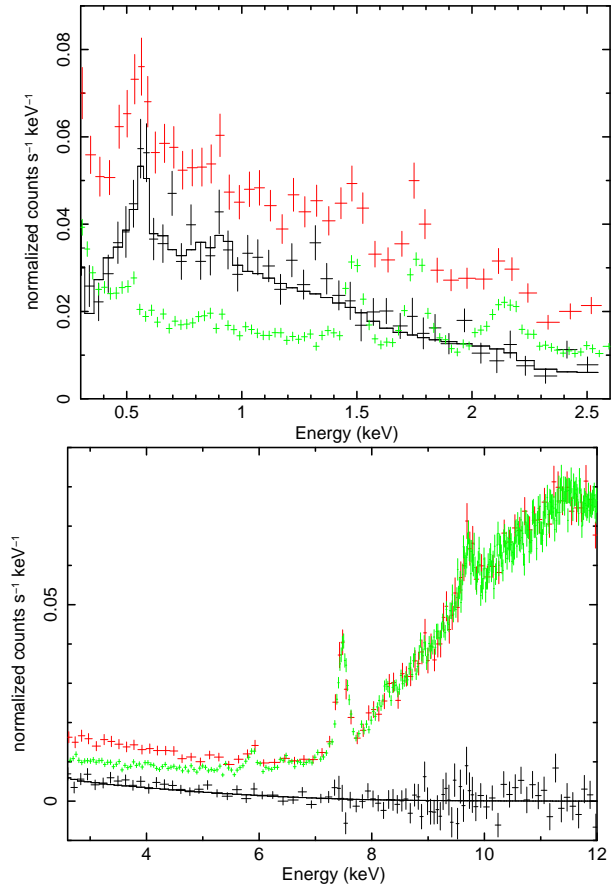

\centering
\includegraphics[scale=0.33,angle=-90]{fig2a.eps}
\hfil
\centering
\includegraphics[scale=0.33,angle=-90]{fig2b.eps}
\hfil
\centering
\caption{Same as Figure 1 for the XIS1 spectrum.}
\end{figure}

Any intrinsic X-ray continuum is indeed negligible. With a maximum
circular velocity $<30$ km s$^{-1}$ \citep{s07a}, X-ray emitting gas
cannot be bound to the Ursa Minor dwarf spheroidal. We may estimate
the emission from unresolved stellar X-ray sources by scaling to the
optical luminosity of Ursa Minor ($\sim 3\times 10^5$ $L_{\odot}$)
using the relation and X-ray spectral parameters in \cite{r08}. There
may also be an intermediate mass black hole in Ursa Minor. Based on
extensions of the observed correlations of nuclear black hole mass
with velocity dispersion \citep{g00,ff05,gra08a,gra08b}, bulge
luminosity or mass \citep{mh03,hr04,ff05}, or dark matter halo mass
\citep{ff05} in early type stellar systems, one expects $M_{\rm
  bh}\sim 300-10^4$ M$_{\odot}$ -- although these correlations may
change or break down at low mass \citep{ghb}. This mass range is
estimated by using the measured Ursa Minor velocity dispersion
\citep{w04} and by considering stellar mass-to-light ratios of $\sim
1-3$ \citep{zgz} and a dark matter virial mass $\sim 10^9$
$M_{\odot}$. \cite{mft} estimate an expected X-ray luminosity of
$7\times 10^{34}$ erg s$^{-1}$ for $M_{\rm bh}=2.3\times 10^4$
$M_{\odot}$. We plot the {\it Suzaku} spectrum for an index 1.7 power
law with $L_X =3\times 10^{34}$ erg s$^{-1}$, and an optimistic
estimate of the unresolved stellar emission, along with the
background-subtracted XIS0+3 spectrum of Ursa Minor in Figure 3.

\begin{figure}
\includegraphics[angle=-90,scale=.33]{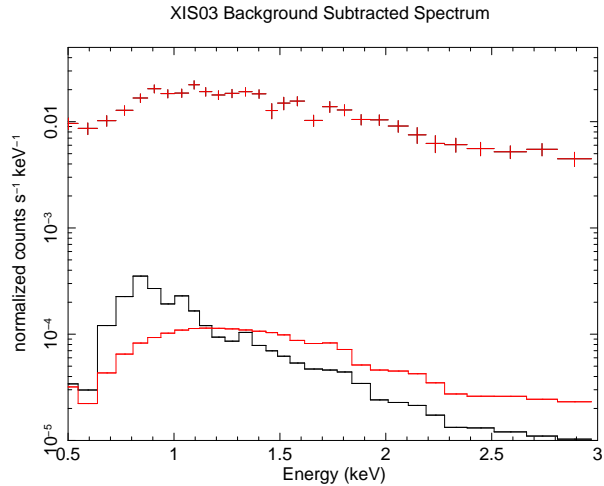}
\caption{Background-subtracted XIS0+3 spectrum of Ursa Minor (crosses)
compared with optimistic estimates of intrinsic X-ray emission from
unresolved stellar sources (black histogram) and an AGN associate
with an intermediate mass ($M_{\rm bh}=10^4$ $M_{\odot}$) black hole (red
histogram.)}
\end{figure}

Typically, in {\it Suzaku} extended source spectral analysis
\citep{mat07,tak07,taw08,tok08, s08}, the estimated NXB is directly
subtracted and the source model often extended to include the GXB and
CXB (perhaps with a subset of GXB/CXB parameters fixed based on
previous observations). Offset or ``blank-sky'' fields may be employed
to directly subtract or fix the parameters of these components. The
NXB-subtracted spectrum is binned and best fits and confidence levels
are derived from $\chi^2$ statistics. Many previous limits on sterile
neutrino radiative decay line fluxes are derived employing variations
on these procedures, adding an ``extra'' line component and computing
the line flux that results in an increase in $\chi^2$ by some amount
corresponding to the desired confidence level. The appropriate
$\Delta\chi^2$ may be chosen under the assumption of Gaussian
statistics (in which case $\Delta\chi^2$ is distributed as $\chi_q^2$,
where $q$ is the number of parameters of interest) or based on
simulations \citep{a07}. \cite{pro02} and others have identified
shortcomings of this general approach, although some may be obviated
by carrying out simulations.

We developed and applied a carefully considered alternative approach
that we believe is statistically well founded. We do not subtract the
NXB and use the unbinned data that is more appropriate for searches
for a narrow, weak spectral line. Spectral fits must thus include an
NXB component. Best fits are determined through minimization of the
maximum likelihood Cash statistic (C-statistic; Cash 1979), as
appropriate for unbinned and unsubtracted data that obeys Poisson
statistics, using version 12.4 of the XSPEC spectral fitting
package\footnote{http://heasarc.gsfc.nasa.gov/docs/xanadu/xspec/}.

A baseline model is constructed as follows. (1) We fit the generated
NXB spectra (Section 2.2.2) over the full 0.4-12 keV band with a model
consisting of two power-law components plus 10 narrow Gaussian lines
with energies fixed at values given in the {\it Suzaku} technical
description\footnote{http://heasarc.gsfc.nasa.gov/docs/suzaku/prop\_tools\\/suzaku\_td/suzaku\_td.html}. XIS0+3
and XIS1 spectra are fitted independently at this stage. (2) Next, we
partition the bandpass into five intervals (0.45-0.65 keV, 0.65-1.3
keV, 1.3-2.6 keV, 2.6-5.2 keV, 5.2-E$_{max}$ keV, where E$_{max}$
=10.4 keV for XIS0+3 and 7.6 keV for XIS1) and refit each NXB spectral
segment with the model described above (including only those lines
that fall within the segment under consideration), allowing line
energies and normalizations to freely vary. If fits are improved by
the inclusion of additional lines, by the replacement of Gaussian with
Lorentzian line profiles, or by introducing finite line widths, these
adjustments are implemented.  The partitioning results in two
line-free (0.65-1.3 keV, 2.6-5.2 keV) and two line-rich segments
(1.3-2.6 keV, 5.2-E$_{max}$ keV) with total counts fairly evenly
apportioned among these four (see Table 1 for total bandpass counts
for each detector). (3) All 10 XIS0+3 and XIS1 NXB sub-spectra are
then simultaneously fitted with line energies fixed at their locally
determined values and one of the two power laws constrained to be the
same for all spectra. This provides a phenomenological baseline NXB
model. A baseline GXB+CXB model consists of an unabsorbed thermal
plasma ({\bf apec}) model, and variable abundance plasma ({\bf vapec})
and power-law ({\bf plaw}) models absorbed by the Galactic column
density in the direction of Ursa Minor, as described in the previous
section. The baseline NXB and GXB+CXB models are concatenated to form
the total baseline model that we use as a starting point for our
ultimate best fits. The best-fit {\bf vapec} component require a
sub-solar oxygen abundance -- all other elements are consistent with
solar abundances and fixed as such.

\begin{deluxetable}{lrc}
\tablewidth{0pt}
\tablecaption{Exposures Times and Spectral Counts}
\tablehead{
\colhead{Detector} & \colhead{Time} & \colhead{Counts}}
\startdata 
XIS0+3 & 138900 & 14170\\ 
XIS1 & 69450 & 10610\\ \enddata 
\end{deluxetable}

The total Ursa Minor spectra are likewise partitioned into 10 segments
(five each for XIS0+3 and XIS1). The baseline model is applied and
then new simultaneous fits to the 10 spectra are found by minimizing
the C-statistic. All NXB line energies and one NXB power-law component
spectral index are now fixed; all constrainable parameters for the GXB
(two temperatures, one oxygen abundance, two normalizations) and CXB
(one power-law index, one normalization) components may freely vary
but are tied for all 10 spectral segments since they are best
determined globally. GXB+CXB model components differ only by a
constant multiplicative scale factor for XIS1 relative to XIS0+3, to
allow for residual flux cross-calibration offsets.  All NXB line
fluxes, one power-law index and two power-law norms are allowed to
independently vary in each spectral segment for each detector
(allowing both power-law indices to vary independently did not improve
the fits). Backfitting the best-fit GXB+CXB model components to the
NXB-subtracted spectra described above yields acceptable fits;
however, this is not the case when fitting the estimated NXB spectra
with the best-fit NXB components. Evidently there are non-negligible
(for our purposes) differences between the estimated and true
NXB. These are apparent as residuals above 4 keV in the NXB-subtracted
spectra (these residuals do not ruin global fits because of the
relatively small number of $>4$ keV source counts), and introduce
systematic effects in the standard approach that we automatically
account for in our analysis. When considering the inclusion of an
additional (sterile neutrino) emission line, allowing the full range
of NXB+GXB+CXB parameters of interest to vary provides a relatively
conservative limit, while also utilizing unbinned, unsubtracted
spectra enables us to consider energies for the extra line up to $\sim
10$ keV (in XIS0+3). The issue of placing upper confidence limits on a
quantity bounded from below \citep{fc98,pro02} is mitigated since we
can allow line fluxes to go negative without the total spectrum
falling below zero.\footnote{The ``slice method'', wherein the total
  spectrum is partitioned into slices corresponding to the appropriate
  energy resolution, and the total counts in each slice used to derive
  an upper limit \citep{rhp} may provide a yet more conservative
  limit. This method, unnecessarily in our view, discards
  well-established characteristics of known X-ray emission
  components.}

\subsubsection{Spectral Fits}

The best-fit NXB+GXB+CXB model to the Ursa Minor spectrum constitutes
the null hypothesis that the X-ray spectrum consists solely of
background components, to which we wish to test the hypothesis that an
additional narrow\footnote{Dwarf spheroidal dark matter velocity
dispersions, $10-30$ km s$^{-1}$, are much less than the XIS energy
resolution.}  emission line -- presumably from sterile neutrino
radiative decay -- may be present. We emphasize that the spectra are
minimally processed -- screened of spurious events, but extended to
the maximum useful bandpass, and unbinned and unsubtracted.

The final model has 62 parameters as follows: the 8 GXB+CXB parameters
described above (including the XIS1/XIS0+3 multiplicative offset) and
54 NXB parameters -- one power-law index and two power-law
normalizations for each of the 10 spectra, nine emission line
normalizations for each detector, plus two additional lines for XIS0+3
with energies above the XIS1 maximum energy of 7.6 keV, and two line
widths for each detector. The best fit (Figure 4) has a minimized
C-statistic of 2863 for 2879 bins (2817 degrees of freedom); best fit
parameters are displayed in Tables 2 and 3 for GXB+CXB and NXB
parameters, respectively (in the former the XIS1/XIS0+3 factor was
found to be $\approx 1.1$). Parameters are consistent with previous
work \citep{mat07,tak07,s08,w08} within their established variation on
the sky as well as with the best fits to NXB-subtracted spectra of
Ursa Minor described above.

\begin{figure}
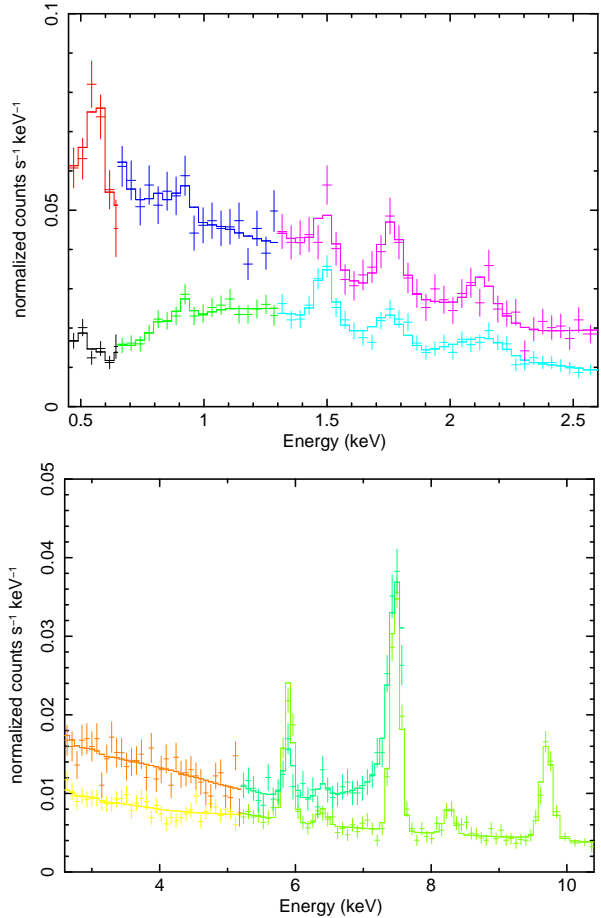

\centering
\includegraphics[scale=0.33,angle=-90]{fig4a.eps}
\hfil
\centering
\includegraphics[scale=0.33,angle=-90]{fig4b.eps}
\hfil
\centering 
\caption{Low-energy ($<2.6$ keV; {\bf above}) and high-energy ($>2.6$
  keV; {\bf below}) XIS0+3 and XIS1 spectra; the XIS1 spectrum has the
  higher count rate. Each spectrum is divided into five segments,
  delineated here by different colors, and all ten segments are
  simultaneously fitted as described in the text. Histograms trace the
  best-fit model.}  
\end{figure}

\begin{deluxetable}{llcl}
\tablewidth{0pt}
\tablecaption{Best Fit CXB+GXB Parameters}
\tablehead{
\colhead{Model} & \colhead{kT or $\Gamma$} & \colhead{O Abundance} &
\colhead{Norm}}
\startdata 
{\bf apec} & 0.075 & \nodata & 0.0058\\
{\bf vapec} & 0.22 & 0.21 & 0.0016\\
{\bf plaw} & 1.45 & \nodata & 0.00087\\
\enddata 

\tablecomments{Temperatures in keV; oxygen abundances relative to
  solar \citep{ag89}; plasma model norms in units of $10^{-14}\int
  DVn_en_h/4\pi d^2$, where $d$ is the source distance in cm and $n_e$
  and $n_h$ are the electron and hydrogen densities in cm$^{-3}$;
  power-law norm in units of photons keV$^{-1}$ cm$^{-2}$ s$^{-1}$ at
  1 keV (corresponding to $6.08\times 10^{-9}$ erg cm$^{-2}$ s$^{-1}$
  over the 2-10 kev band). Norms correspond to a $20'$ circular
  region, that is, 1256.6 sq-arcminute=0.0001063 str.}
\end{deluxetable}

\subsection{Sterile Neutrino Emission Line Flux Upper Limits}

The null hypothesis asserts that the Ursa Minor X-ray spectrum
consists solely of background (NXB, GXB, CXB) components, with
estimated parameters corresponding to the minimum value of the
C-statistic ($C_{\rm null}$). We may add a narrow (sterile neutrino
radiative decay) line and independently refit the augmented model to
obtain $C_{\rm line}(E_{\rm line})$ for a range of line energies
$E_{\rm line}$. Given the large number of counts in our unsubtracted
spectra, $\Delta C\equiv C_{\rm null}-C_{\rm line}(E_{\rm line})$ is
ideally expected to be distributed as the $\chi^2$ statistic with one
degree of freedom, and the confidence level that a line is detected by
$P(\chi_1^2<\Delta C)$ \citep{k00}. In the absence of a significant
detection, an $\alpha$-upper limit line flux is that which corresponds
to a value of $\Delta C$ such that $P(\chi_1^2<\Delta C)=\alpha$
\citep{yaq98}.

We perform Monte Carlo spectral simulations in order to test how
accurately these expectations are realized.  The parameters from the
best-fit null hypothesis model to the observed data are used to
generate fake spectra that are re-fitted with the null hypothesis
model. These second generation fits are used as seeds for subsequent
Monte Carlo realizations \citep{but05,mrb06,por07} that are generated
and fitted in a manner that matches the actual spectra and their best
fits (same exposure times and instrument responses, same assignment of
fixed, tied, and free parameters). Fitting these simulated spectra
with the null hypothesis model determines the minimized fit statistic
$C_{\rm null}$. Each simulated data set is additionally re-fitted with
models that include an additional line at an (non-fixed) energy
$E_{\rm line}$ that is stepped through the analysis bandpass (at 25 eV
intervals) to determine the minimum fit statistic $C_{\rm line}(E_{\rm
  line})$ for each $E_{\rm line}$. 1000 simulated spectra are
generated for each hypothesis (null or null-plus-line-at-$E_{\rm
  line}$ for each $E_{\rm line}$). The cumulative probability
distribution $P(<\Delta C)$, where again $\Delta C\equiv C_{\rm
  null}-C_{\rm line}(E_{\rm line})$, yields an experimental estimate
of the corresponding probability of rejecting the null hypothesis. The
upper limit may then be estimated as described above, using the
experimental, instead of $\chi_1^2$ distribution. We find that
$P(<\Delta C)$ matches $P(<\chi_1^2)$ only in spectral intervals with
high total count rates and low NXB (i.e., in NXB-line-free
regions). The joint distribution over all energies is
well-approximated by $P(<\chi_2^2)$, and at no energies does
$P(<\Delta C)$ fall below $P(<\chi_3^2)$ (Figure 5).

\begin{figure}
\plotone{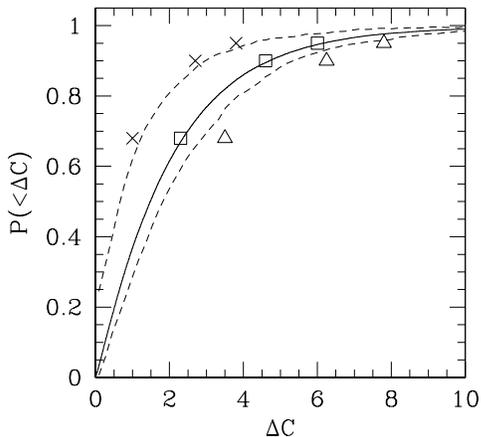}
\caption{Cumulative probability of $\Delta C\equiv C_{\rm null}-C_{\rm
line}$ from simulations. The solid curve shows the sum over all energies,
while the broken curves shows the full range spanned by simulations at
individual energies. Crosses, boxes, and triangles show the 68, 90,
and 95\% cumulative probabilities for the $\chi_1^2$, $\chi_2^2$, and
$\chi_3^2$ distributions, respectively.}
\end{figure}

We adopt $\Delta C=10$ upper limits. This corresponds to $P(<\Delta
C)\sim 0.99$ in the joint distribution; at no energy was $P(\Delta
C<10)$ less than 0.984 (or greater than 0.996). The $\Delta C=10$
upper and lower limits are plotted in Figure 6. Since no lower limit
is $>0$, we infer that there is no line detected at $\ge 99\%$
confidence. To derive limits in the vicinities of NXB lines we fix
their line fluxes (otherwise, in fits and simulations, these are free
to vary) at their best-fit values. The resulting limits are very
nearly equal to the $\Delta C=10$ uncertainty in the NXB line fluxes
and, as they neglect any systematic errors in the NXB line strengths,
will underestimate the upper limit of any additional emission line
with precisely the same energy as the background features. The limits
are tightest in spectral intervals where lines from the NXB and GXB
are sparse or absent (especially $\sim 2-5$ keV). When we compare
these limits to those more conventionally derived from the binned,
background-subtracted spectra, we find that the $\Delta C=10$ limits
roughly correspond to $\Delta\chi^2=7$ in the $<4$ keV region where
the NXB is small or cleanly subtracts out.

\begin{figure}
\plotone{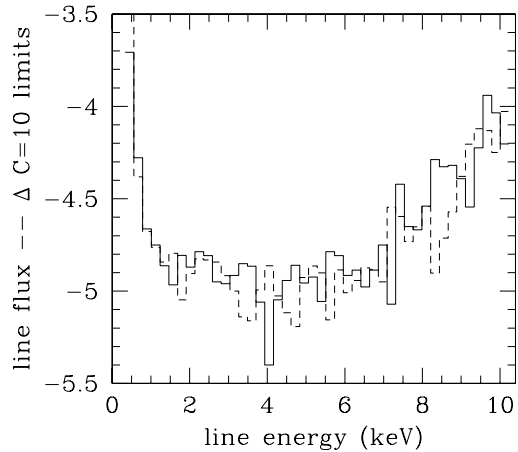}
\caption{Log of $\Delta C=10$ line flux upper limits in photons
cm$^{-2}$ s$^{-1}$ (solid histogram). Also shown is the log of the
absolute value of the lower limits (that are all negative; broken
histogram).}
\label{figure:photon_flux}
\end{figure}

\section{Limits on Sterile Neutrino Parameters}

The integrated line flux from sterile neutrinos decaying
at rate $\Gamma_{\rm st}=10^{-27}\Gamma_{-27}$ s$^{-1}$ from a dark matter
mass in projection of $10^{7}M_{7}$ $M_{\odot}$ at a distance
100$d_{100}$ kpc is
\begin{equation}
{\mathcal F}_{\rm line}=7.5\times
10^{-15}~\Gamma_{-27}f_{\rm st}M_{7}d_{100}^{-2}~{\rm erg~cm}^{-2}~{\rm
  s}^{-1},
\end{equation}
where $f_{\rm st}$ is the dark matter fraction in sterile neutrinos. The
corresponding number flux of $E_\gamma=m_{\rm st}/2$ photons from decays
of particles with mass $m_{\rm st}$ that make up the total mass
$10^{7}M_{7}f_{\rm st}$ $M_{\odot}$ is given by
\begin{eqnarray}
F_{\rm line}& =& 9.4\times 10^{-6}~\Gamma_{-27} \, f_{\rm st} \, M_{7}
d_{100}^{-2} \left(\frac{\rm keV}{m} \right) \\ & = & 5.3\times 10^{-10} \left(\frac{\sin^2
  \theta}{10^{-10}} \right) \left(\frac{m_{\rm st}}{\rm keV} \right)^4\,
f_{\rm st}\, M_{7} d_{100}^{-2} 
\label{photon_flux}
\end{eqnarray}
${\rm cm}^{-2}\ {\rm s}^{-1}$, where we have used the expression for $\Gamma_{\rm st}$ introduced in equation
(1).

One can use this relation to set limits on sterile neutrino mass and
mixing angle using the photon flux limits shown in
Figure~\ref{figure:photon_flux}.  The dark matter halo mass
distribution in Ursa Minor may be estimated using stellar
dynamics. Strigari and collaborators \citep{s07a,s07b,s08a, s08b}
applied maximum likelihood analysis to constrain the parameters of a
generalized Navarro-Frenk-White (NFW) profile \citep{d05}.  Based on
their confidence region for an NFW model (inner dark matter density
slope of $-1$) in \cite{s08a}, and the general integrated constraints
\citep{g07,w07,s07b,pmn08,s08b}, the projected mass of Ursa Minor
within $20'$ ($\sim 0.4$ kpc at our adopted distance of 69 kpc;
Grebel, Gallagher, \& Harbeck 2003) is well determined from the
projection of the NFW model \citep{b96,gk02,ymk03} to be $\sim 6\times
10^7$ $M_{\odot}$. This is consistent with the weighted average
derived from the estimates in the papers quoted above (Figure 7; these
adopt various assumptions about the cuspiness of the profile) that has
a variance of $\sim 25$\%, although the range of models formally
allowed in \cite{s08a} spans factors of $\sim 2$ in either direction
at that radius. The corresponding excluded dark matter decay rate
$\Gamma_{\rm st}$ (for $f_{\rm st}=1$), derived from the $\Delta C=10$
line flux upper limits, is shown in Figure 8 for line energies
$0.45<\epsilon_{\rm keV}<10.4$.

\begin{figure}
\plotone{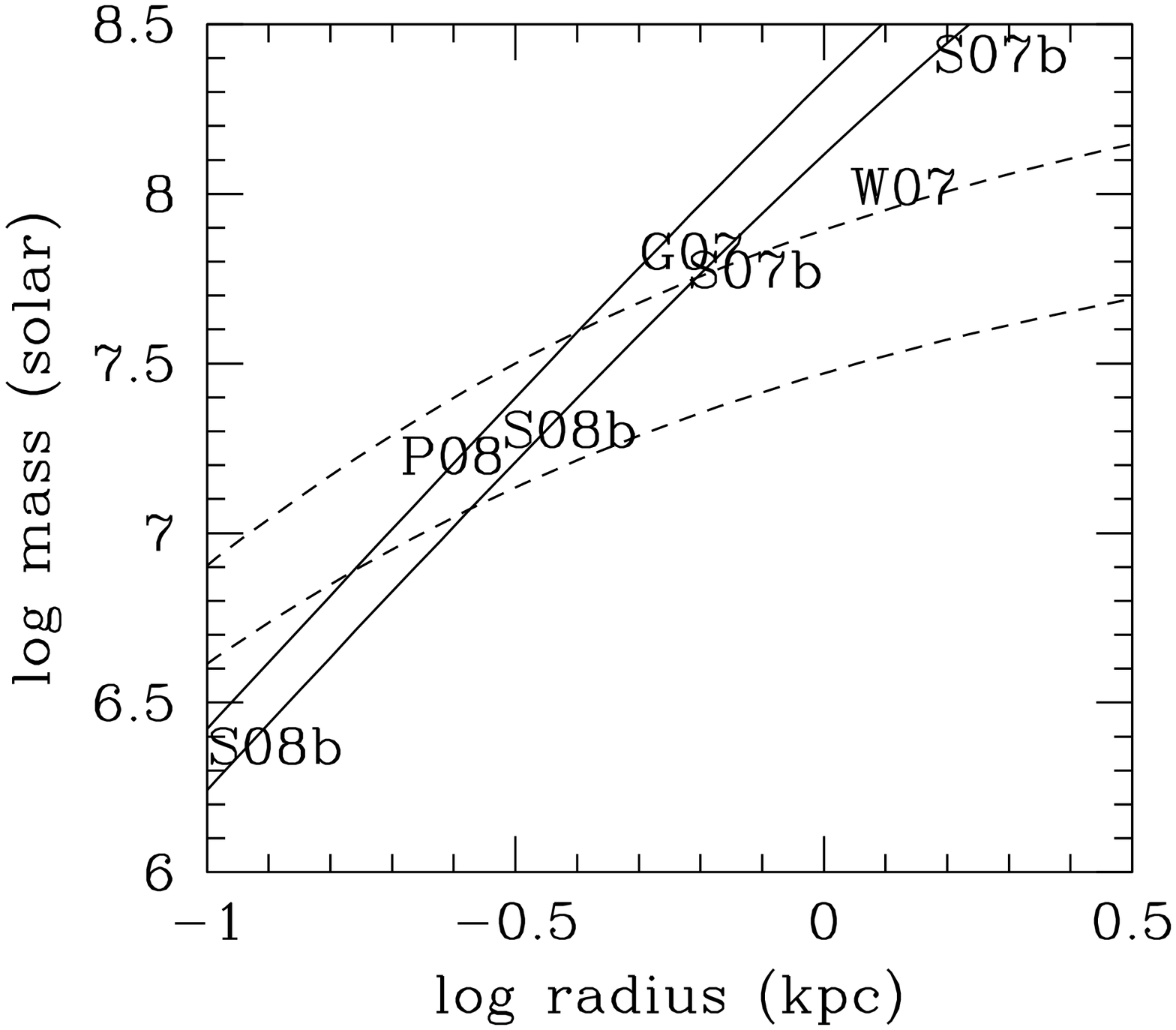}
\caption{Solid (broken) curve shows the 90\% confidence limits of
the least (most) compact mass profile from \cite{s08a}.  Also shown
are estimates, at various radii from \cite{g07} (G07), \cite{w07}
(W07), \cite{s07b} (S07b), \cite{pmn08} (P08), and \cite{s08b}
(S08b).}
\label{figure:mass_profile}
\end{figure}

If one assumes that all the relic sterile neutrinos are produced via
the DW mechanism, the absence of line emission from Ursa Minor
provides an upper limit on the mixing angle as a function of mass that
is independent of whether there are additional production mechanisms
(which could only increase the abundance).  The flux from
equation~(\ref{photon_flux}), calculated for a given value of $f_{\rm
st}$, should not exceed the photon flux shown in
Figure~\ref{figure:photon_flux}.  In a given cosmological scenario, if
the sterile neutrino relic abundance is a monotonic function of mass
and mixing angle, $f_{\rm st}=p(\sin^2\theta,m_{\rm st})$, then the
region allowed in the $\sin^2\theta-m_{\rm st}$ plane is given by
\begin{equation}
\sin^2\theta_{max,p}=\min(\sin^2\theta_{max,1},\sin^2\theta_{max,2}),
\label{limits_gen}
\end{equation}
where $\sin^2\theta_{max,1}$ is determined from
$f_{\rm st}=p(\sin^2\theta,m_{\rm st})<1$ and $\sin^2\theta_{max,2} $ from
the requirement that the flux in equation~(\ref{photon_flux}) not
exceed the observed flux shown in Figure~6.

We derive two different limits that provide answers to the following
questions: (1) whether the existence of a sterile neutrino with a
given mass and mixing angle is consistent with standard cosmological
history and (2) whether sterile neutrinos of a given mass and mixing
angle can account for 100\% of dark matter.  These two constraints
are, of course, different.

For the first limit, we take $f_{\rm st}=p_{\rm DW}(m_{\rm
st},\theta)$ to be a function of mass and mixing angle and we assume
the DW production mechanism, which gives the minimal abundance of
sterile neutrinos in standard Big Bang cosmology.  \cite{asl} provided
fitting formulae for sterile neutrino abundance produced by the DW
mechanism of the form
\begin{equation}
p_{\rm DW}(\sin^2\theta,m_{\rm st})=\alpha
\left(\frac{\sin^2\theta}{10^{-10}}\right) \left(\frac{m_{\rm
st}}{1\,\rm keV}\right)^{\beta},
\end{equation}
as well as pairs $(\alpha,\beta)$ for average, minimum, and maximum
production that account for hadronic uncertainties.  The excluded
region in the $m_{\rm st}-\theta$ plane, for production dominated by
the DW mechanism, is given by equation (4), with
\begin{equation}
\left(\frac{\sin^2\theta_{max,1}}{10^{-10}}\right)=\alpha^{-1}
\left(\frac{m_{\rm st}}{1\,\rm keV}\right)^{-\beta},
\end{equation}
and
\begin{equation}
\sin^2\theta_{max,2}=(\sin^2\theta_{max,1}\times\sin^2\theta_{max,0}[1])^{1/2},
\end{equation}
where $\sin^2\theta_{max,0}[1]$ is determined by setting $f_{\rm
st}=1$. The resulting exclusion region is shown in
Figure~\ref{figure:Suzaku_limits} for the minimal rate of DW
production consistent with the results of \cite{asl}, which provides
the most conservative constraints.  This region is excluded regardless
of the physics responsible for mass generation of neutrinos or any
other physics beyond direct mixing between sterile and active
neutrinos.

For the second kind of limit, we determine the part of parameter space
for which sterile neutrinos can account for all the cosmological dark
matter, while still being consistent with {\it Suzaku} observations.
Here we set $f_{\rm st}=1$ without reference to any specific
production mechanism.  The excluded region (for $f_{\rm st}=1$)
corresponding to the $\Delta C=10$ upper limits on $\Gamma_{\rm
st}f_{\rm st}$ shown in Figure 8 is delineated by the solid line in
Figure~\ref{figure:Suzaku_limits}. Sterile neutrinos occupying the
parameter space to the right of the solid line cannot make up all of
the dark matter (although such a particle may exist in nature, in
contrast with the previous limit).

\begin{figure}
\plotone{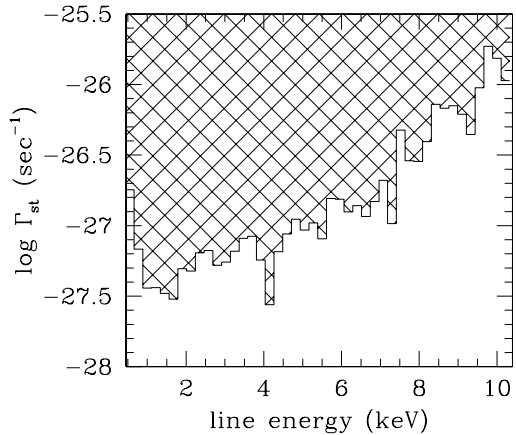}
\caption{Excluded dark matter radiative decay rates (hatched region)
as a function of energy averaged over the inner 0.4 kpc of Ursa Minor,
assuming that 100\% of the dark matter is composed of sterile neutrinos
($f_{\rm st}=1$).}
\end{figure}

\begin{figure}
\plotone{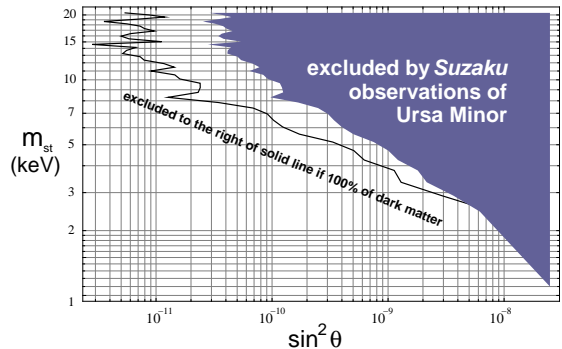}
\caption{Sterile neutrino parameter space to the right of the
  solid curve is excluded by the {\it Suzaku} observation of Ursa
  Minor if dark matter is solely composed of sterile neutrinos
  produced by some (unspecified) mechanism.  The solid exclusion
  region is model-independent, based only on the assumption of the
  standard cosmological history below the temperature of a few hundred
  MeV, when the DW production by neutrino oscillations takes place.  }
\label{figure:Suzaku_limits}
\end{figure}

One can ask which dark-matter particle mass in the form of sterile
neutrinos, for $f_{\rm st}=1$, produced solely by the DW mechanism is
consistent with {\it Suzaku} observations.  Here one can set an upper
limit on the sterile neutrino mass: $m_{\rm st}<2.5$~keV.  However,
the DW mechanism with $f_{\rm st}=0.1$ is not ruled out at any energy
covered by the {\it Suzaku} spectra ($\sim 1-20$ keV). This is shown
in Figure 10, where our line flux upper limits are compared with the DW
predictions.

\begin{figure}
\plotone{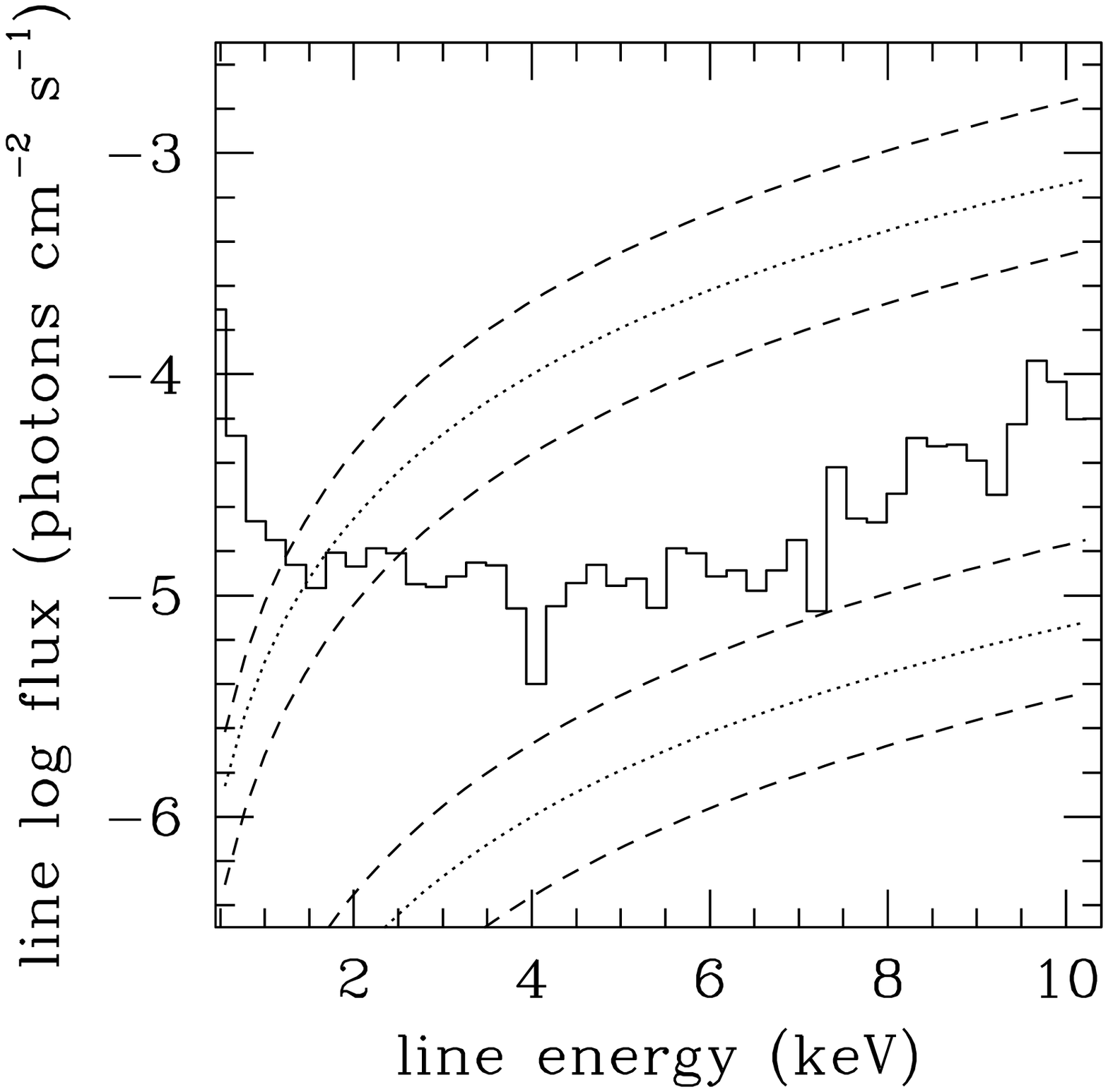}
\caption{Line flux upper limits from Figure 6 re-plotted with the
$f_{\rm st}=1$ (upper curves) and $f_{\rm st}=0.1$ (lower curves) DW
predictions for maximum, average, and minimum production \citep{asl}.}
\end{figure}

\subsection{Comparison to Previous Work}

Because different criteria for estimating line flux upper limits are
used, a precise comparison with previous results is difficult. Our
limits are about a factor of 3 better than those for Ursa Minor in
\cite{bnr} over the overlapping bandpass. The constraints shown in
Figure 9 are comparable to the combined constraints from the Milky Way
\citep{bnr} and M31 \citep{birs}, and slightly tighter than those from
the unresolved {\it Chandra} CXB \citep{a07} and from the ``bullet''
cluster, 1E 0657-56 \citep{brm}. \cite{asl} compiled and plotted
various X-ray limits on the combination of $m_{\rm st}$ and $\theta$.
We attain a comparable level of sensitivity over the entire 1-20 keV
mass range with a {\it Suzaku} observation of a single object.

\section{Discussion, Conclusions, and Future Prospects}

With a single {\it Suzaku} observation of the Ursa Minor dwarf
spheroidal, we have derived limits on the mass and mixing angle of the
sterile neutrino dark matter candidate that are comparable to or better
than previous constraints. These limits are illustrated in
Figure~\ref{figure:Suzaku_limits} for the case where sterile neutrinos
comprise all of the dark matter regardless of how they are produced in
the early universe and, in the more general case, when their abundance
is not assumed, but is calculated under the assumptions that minimize
it. The best estimate of the total projected dark matter mass within
400 pc of $6\times 10^7$ $M_{\odot}$ is adopted here. If solely
produced by the DW mechanism, sterile neutrinos cannot constitute
100\% of the dark matter in Ursa Minor unless they are less massive
than 2.5 keV.

Because of its orbit, {\it Suzaku} has a lower and more stable
background than either the {\it Chandra} or {\it XMM-Newton}
Observatories, and its spectrometers have sharper energy resolution.
Our approach in deriving limits utilizes the C-statistic to
evaluate a very general model fit to the data with minimal
pre-processing and no background subtraction. Thus, there are no a
priori assumptions about the accuracy of the NXB estimate that we
only use to set line energies and shapes, or whether an extra line
feature lies above or below the total continuum. Our final limits are
those that correspond to a change in the C-statistic that occurred in
1\% of our Monte Carlo simulations. These factors may account for why
our constraints represent only modest improvements over previously
published limits despite the more optimal characteristics of {\it
Suzaku}. As the NXB characterization improves, it may be possible to
tighten upper limits in the $m_{\rm st}>2$ keV regime; at lower masses,
the ubiquity of the GXB will remain a barrier until significantly
higher energy resolution is attained -- e.g., with {\em
Astro-H}\footnote{http://www.isas.ac.jp/e/enterp/missions/astro-h/\\index.shtml}.

Since there is some uncertainty as to the dynamical state of dwarf
spheroidals on an individual basis, X-ray constraints obtained from
additional systems are warranted (e.g., our paper -- in preparation --
on the {\it Suzaku} observation of the Draco system). The Sloan
Digital Sky Survey (SDSS) has approximately doubled the number of
galaxies identified with the local group population and revealed a new
sub-population of faint Milky Way satellites \citep{bel07}, including
some with very high mass-to-light ratio \citep{s08a,s08b} that -- with
the right observational setup -- may provide the best targets for
investigations of this type.

Finally, predictions of the Ly$\alpha$ forest power spectrum in fully
self-consistent large-scale structure simulations with sterile
neutrinos can provide complementary constraints in the form of a lower
limit on the particle mass.

Sterile neutrinos remain a viable dark matter candidate. For example,
a sterile neutrino with a mass of $m_{\rm st}\sim 4$ keV and
$\sin^2\theta\sim 4\times 10^{-10}$ could account for all of the dark
matter (or $\sim 15$\%$-75$\% if produced solely by the DW mechanism)
and explain pulsar kicks \citep{kmm} without violating the line flux
upper limits.

\acknowledgments{This research has made use of data obtained from the
  Suzaku satellite, a collaborative mission between the space agencies
  of Japan (JAXA) and the USA (NASA). ML thanks R. Mushotzky for
  providing the initial impetus for this project; U. Hwang, K.
  Hamaguchi, and A. Markowitz for advice and assistance with data
  analysis; he dedicates this paper to the memory of E. Boldt for
  many years of collaboration and conversation on the connections
  between particle physics and astrophysics. The work of ML was
  supported by NASA grants NNXO8AD96G and NNX07AO94G.  The work of AK
  was supported in part by DOE grant DE-FG03-91ER40662 and by NASA
  ATFP grant NNX08AL48G. Support for PLB has come from the AUGER
  membership and theory grant 05 CU 5PD 1/2 via DESY/BMBF and VIHKOS.}

\clearpage

\begin{deluxetable}{lllllllllllr}
\tabletypesize{\scriptsize}
\tablewidth{0pt}
\tablecaption{NXB Line Energies, Best-fit Fluxes, and Flux Upper Limits}
\tablehead{
\colhead{} & \colhead{O K} & \colhead{Al K} & \colhead{Si K} &
\colhead{Au M} & \colhead{?}  & \colhead{Mn K$\alpha$} & \colhead{?} &
\colhead{Mn K$\beta$} & \colhead{Ni K$\alpha$} & \colhead{Ni K$\beta$}
& \colhead{Au L$\alpha$} }
\startdata 
& & & & & & XIS0+3  & & & & &\\
$E_{\rm line}$ & 0.525 & 1.49 & 1.74 & 2.12 & 2.23\tablenotemark{b}
& 5.90 & 6.4 & 6.49 & 7.49 & 8.27 & 9.71\tablenotemark{ab}\\
$F_{\rm line}$ & 0.0 & 3.1 & 0.50 & 0.26 & 20 & 9.7 & 1.3 & 0.04 & 31
& 5.4 & 51 \\
$F_{\rm line,up}$ & 8.19 & 4.84 & 1.96 & 2.18 & 46.6 & 11.8 & 2.79 &
1.72 & 35.4 & 8.82 & 62.4 \\
& & & & & &  XIS1 & & & & &\\
$E_{\rm line}$ & 0.525 & 1.49 & 1.77 & 2.12 &
2.23\tablenotemark{ab} & 5.90 & 6.4 & 6.49 & 7.48\tablenotemark{ab}
& &\\
$F_{\rm line}$ & 3.3 & 2.9 & 3.7 & 2.3 & 0.1 & 4.3 & 1.4 & 0.0 & 72 &
&\\
$F_{\rm line, up}$ & 20.4 & 5.95 & 6.83 & 4.78 & 1840 & 7.52 & 4.69 &
3.76 & 111 & &\\
\enddata 
\tablecomments{identifications, when available, from Table 7.2
of ``The Suzaku Technical Description''
(http://heasarc.gsfc.nasa.gov/docs/suzaku/prop\_tools/suzaku\_td/suzaku\_td.html);
line energies in keV, lines fluxes in $10^{-5}$ photon cm$^{-2}$~s$^{-1}$, 
upper limits correspond to $\Delta C=10$.}
\tablenotetext{a}{~Lorentzian} 
\tablenotetext{b}{~finite width}
\end{deluxetable}


\begin{thebibliography}{}
\bibitem[Abazajian(2006a)]{a06a}
Abazajian, K. N. 2006a, \prd, 73, 063506
\bibitem[Abazajian(2006b)]{a06b}
Abazajian, K. N. 2006b, \prd, 73, 063513
\bibitem[Abazajian, Fuller, \&  Patel(2001)]{afp}
Abazajian, K., Fuller, G. M., \&  Patel, M. 2001, \prd, 64, 023501 
\bibitem[Abazajian, Fuller, \&  Tucker(2001)]{aft}
Abazajian, K., Fuller, G. M., \&  Ticker, W. H. 2001, \apj, 562, 593
\bibitem[Abazajian \& Koushiappas(2006)]{ak06}
Abazajian, K., \& Koushiappas, S. M. 2006, \prd, 74, 023527 
\bibitem[Abazajian et al.(2007)]{a07}
Abazajian, K. N., Markevitch, M., Koushiappas, S. M., \& Hickox,
R. C. 2007, \prd, 75, 063511
\bibitem[Anders \& Grevesse(1989)]{ag89}
Anders, E., \& Grevesse, N. 1989, Geochimica et Cosmochimica Acta 53, 197 
\bibitem[Asaka, Laine, \& Shaposhnikov(2007)]{asl}
Asaka, T., Shaposhnikov, M., \& Laine, M. 2007, J. High Energy Phys., 01, 091
\bibitem[Barger et al.(1995)]{barger} Barger, V., Phillips, 
R.~J.~N., \& Sarkar, S.\ 1995, Physics Letters B, 356, 617 
\bibitem[Bartelmann(1996)]{b96} 
Bartelmann, M. 1996, \aap, 313, 697
\bibitem[Belokurov et al.(2007)]{bel07} 
Belokurov, V., et al. 2007, \apj, 654, 897
\bibitem[Biermann \& Kusenko(2006)]{bk06}
Biermann, P. L., \& Kusenko, A. 2006, Phys. Rev. Lett., 96, 091301
\bibitem[Boyanovsky(2008)]{b08}
Boyanovsky, D. 2008 Phys. Rev. D, 78, 103505
\bibitem[Boyarsky et al.(2008)]{birs}
Boyarsky, A., Iakubovskyi, D., Ruchayskiy, O., \& Savchenko, V. 2008,
\mnras, 387, 1361
\bibitem[Boyarsky et al.(2009)]{blrv}
Boyarsky, A., Lesgourgues, J., Ruchayskiy, O., \& Viel, M. 2009,
J. Cosmology Astropart. Phys., 5, 012
\bibitem[Boyarsky et al.(2006a)]{b06a}
Boyarsky, A., Neronov, A., Ruchayskiy, O., \& Shaposhnikov, M. 2006a,
\mnras, 370, 213
\bibitem[Boyarsky et al.(2006b)]{b06b}
Boyarsky, A., Neronov, A., Ruchayskiy, O., \& Shaposhnikov, M. 2006b,
\prd, 74, 103506
\bibitem[Boyarsky et al.(2006c)]{b06c} Boyarsky, A., Neronov, A.,
Ruchayskiy, O., Shaposhnikov, M., \& Tkaachev, I. 2006c, Phys. Rev
Lett., 97, 261302
\bibitem[Boyarsky, Nevalainen, \& Ruchayskiy(2007)]{bnr}
Boyarsky, A., Nevalainen, J., \& Ruchayskiy, O. 2007, \aap, 471, 51
\bibitem[Boyarsky, Ruchayskiy, \& Markevitch(2008)]{brm}
Boyarsky, A., Ruchayskiy, O., \& Markevitch, M. 2008. \apj, 673, 752
\bibitem[Butler et al.(2005)]{but05}
Butler, N., Ricker, G., Vanderspek, R., Ford, P., Crew, G., Lamb,
D. Q., \& Jernigan, J. G. 2005, \apj, 627, L9
\bibitem[Cash(1979)]{c79}
Cash, W. 1979, \apj, 228, 939
\bibitem[Diemand et al.(2005)]{d05} 
Diemand, J., Zemp, M., Moore, B., Stadel, J., \& Carollo, C. 2005,
\mnras, 364, 665
\bibitem[Dodelson \& Widrow(1994)]{dw94}
Dodelson, S., \& Widrow, L. M. 1994, \prl, 72, 17
\bibitem[Dolgov(2002)]{d02}
Dolgov, A. D. 2002, Phys. Rept., 370, 333
\bibitem[Dolgov \& Hansen(2002)]{dh} Dolgov, A.~D., \& Hansen, S.~H.\ 2002, Astroparticle Physics, 16, 339  
\bibitem[Feldman \& Cousins(1998)]{fc98}
Feldman, G. J., \& Cousins, R. D. 1998, \prd, 57, 3873
\bibitem[Ferrarese \& Ford(2005)]{ff05}
Ferrarese, L., \& Ford, H. 2005, \ssr, 116, 523
\bibitem[Fuller et al.(2003)]{Fuller03} Fuller, G.~M., Kusenko, 
A., Mocioiu, I., \& Pascoli, S.\ 2003, \prd, 68, 103002 
\bibitem[Fryer \& Kusenko(2006)]{Fryer} 
Fryer, C.~L., \& Kusenko, A.\ 2006, \apjs, 163, 335 
\bibitem[Gao \& Theuns(2007)]{gao} Gao, L., \& Theuns, T.\ 2007, Science, 317, 1527 
\bibitem[Gebhardt et al.(2000)]{g00} 
Gebhardt, K., et al. 2000, \apj, 539, 13
\bibitem[Gilmore et al.(2007)]{g07} 
Gilmore, G., Wilkinson, M. I., Wyse, R. F. G., Kleyna, J. T., Koch,
A., Evans, N. Wyn, \& Grebel, E. K. 2007, \apj, 663, 948
\bibitem[Graham(2008a)]{gra08a} 
Graham, A. W. 2008a, \apj, 680, 143
\bibitem[Graham(2008b)]{gra08b} 
Graham, A. W. 2008b, PASA, 25, 167
\bibitem[Greene, Ho, \& Barth(2008)]{ghb} 
Greene, J. E., Ho, L. C., \& Barth, A. J. 2008, \apj, 688, 159
\bibitem[Golse \& Kneib(2002)]{gk02} 
Golse, G., \& Kneib, J.-P. 2002, \aap, 390, 821
\bibitem[Grebel, Gallagher, \& Harbeck(2003)]{ggh03} 
Grebel, E. K., Gallagher, J. S., III, \& Harbeck, D. 2003, \aj, 125, 1926
\bibitem[H\"aring \& Rix(2003)]{hr04} 
H\"aring, N., \& Rix, H.-W. 2004, \apj, 604, L89
\bibitem[Hidaka \& Fuller(2007)]{hf07}
Hidaka, J. \& Fuller, G. M. 2007, \prd, 76, 083516 
\bibitem[Ishisaki et al.(2007)]{i07}
Ishisaki, Y., et al. 2007, \pasj, 59, S113
\bibitem[Kadota(2008)]{kadota} Kadota, K.\ 2008, \prd, 77, 063509 
\bibitem[Koyama et al.(2007)]{k07}
Koyama, K., et al. 2007, \pasj, 59, S23
\bibitem[Kurczynski et al.(2000)]{k00}
Kurczynski, P., et al  2000, \apj, 543, 77
\bibitem[Kusenko(2004)]{k04}
Kusenko, A. 2004, J. Mod. Phys. D, 13, 2065
\bibitem[Kusenko(2006)]{k06}
Kusenko, A. 2006, \prl, 97, 241301
\bibitem[Kusenko, Mandal, \& Mukherjee(2008)]{kmm}
Kusenko, A., Mandal, B. P., \& Mukherjee, A. 2008, \prd, 77, 123009
\bibitem[Kusenko \& Segr\`e(1997)]{ks97}
Kusenko, A., \& Segr\`e, G. 1997, Phys.\ Lett.\ B, 396, 197
\bibitem[Laine \& Shaposhnikov(2008)]{ls08}
Laine, M., \& Shaposhnikov, M. 2008, JCAP, 06, 31
\bibitem[Maccarone, Fender, \& Tzioumis(2005)]{mft} 
Maccarone, T. J., Fender, R. P., \& Tzioumis, A. K. 2005, \mnras, 356, 17
\bibitem[Marconi \& Hunt(2003)]{mh03} 
 Marconi A., \& Hunt, L. K. 2003, \apj, 589, L21
\bibitem[Markowitz et al.(2006)]{mrb06} 
Markowitz, A., Reeves, J. N., \& Braito, V. 2006, \apj, 646, 783
\bibitem[Matsushita et al.(2007)]{mat07}
Matsushita, K., et al. 2007, \pasj, 59, S327
\bibitem[Meiksin(2009)]{reviewLya} 
Meiksin, A.~A.\ 2009, Rev. Mod. Phys., in press
\bibitem[Mitsuda et al.(2007)]{m07}
Mitsuda, K. et al. 2007, \pasj, 59, S1
\bibitem[Munyaneza \& Biermann(2005)]{mb05} Munyaneza, F., \& Biermann, P.~L.\ 2005, \aap, 436, 805 
\bibitem[Munyaneza \& Biermann(2006)]{mb06} Munyaneza, F., \& Biermann, P.~L.\ 2006, \aap, 458, L9 
\bibitem[Nakajima et al.(2008)]{n08}
Nakajima, H., et al. 2008, \pasj, 60, S1
\bibitem[Narayanan et al.(2000)]{n00}
Narayanan, V. K., Spergel, D. N., Dav\`e, R., \& Ma, C.-P. 2000, \apj, 543, L103
\bibitem[Pal \& Wolfenstein(1982)]{pw} 
Pal, P.~B., \& Wolfenstein, L.\ 1982, \prd, 25, 766 
\bibitem[Palazzo et al.(2007)]{p07} 
Palazzo, A., Cumberbatch, D., Slosar, A., \& Silk, J. 2007, \prd, 76, 103511
\bibitem[Penarrubia, McConnachie, \& Navarro(2008)]{pmn08} 
Penarrubia, J., McConnachie, A. W., \& Navarro, J. F. 2008, \apj, 672, 904
\bibitem[Petraki(2008)]{pet08}
Petraki, K. 2008, \prd, 77, 105004
\bibitem[Petraki \& Kusenko(2008)]{pk08}
Petraki, K., \& Kusenko, A. 2008, \prd, 77, 065014
\bibitem[Porquet et al.(2007)]{por07} 
Porquet, D., et al. 2007, \aap, 473, 67
\bibitem[Protassov et al.(2002)]{pro02} 
Protassov, R., van Dyk, D. A., Connors, A., Kashyap, V. L., \&
Siemiginowska, A. 2002, \apj, 571, 545
\bibitem[Revnivtsev et al.(2008)]{r08} 
Revnivtsev, M., Churazov, E., Sazonov, S., Forman, W., \& Jones,
C. 2008, \aap, 490, 37
\bibitem[Riemer-S\o{}rensen, Hansen, \& Pedersen(2006)]{rhp}
Riemer-S\o{}rensen, S., Hansen, S. H., \& Pedersen, K. 2006, \apj, 644,
L33
\bibitem[Riemer-S\o{}rensen et al.(2007)]{r07}
Riemer-S\o{}rensen, S., Pedersen, K., Hansen, S. H., \& Dahle, H. 2007,
\prd, 76 043524
\bibitem[Sato et al.(2008)]{s08}
Sato, K., et al. 2008, \pasj, 60, S333
\bibitem[Serlemitsos et al.(2007)]{se07}
Serlemitsos, P. J., et al. 2007, \pasj, 59, S9
\bibitem[Shaposhnikov \& Tkachev(2006)]{st} Shaposhnikov, M., \& Tkachev, I.\ 2006, Physics Letters B, 639, 414 
\bibitem[Shi \& Fuller(1999)]{sf99}
Shi, X.D. \& Fuller, G. M. 1999, \prl, 82, 2832
\bibitem[Stasielak, Biermann, \& Kusenko(2007)]{s07}
Stasielak, J., Biermann, P. L., \& Kusenko 2007, ApJ, 654, 290
\bibitem[Strigari et al.(2007a)]{s07a} 
Strigari, L. E., Bullock, J. S., \& Kaplinghat, M., Diemand, J., Kuhlen,
M., \& Madau, P. 2007, \apj, 669, 676
\bibitem[Strigari et al.(2007b)]{s07b}
Strigari, L. E., Koushiappas, S. M., Bullock, J. S., \& Kaplinghat,
M. 2007, \prd, 75, 3526
\bibitem[Strigari et al.(2008a)]{s08a} 
Strigari, L. E., Koushiappas, S. M., Bullock, J. S., Kaplinghat,
M., Simon, J. D, Geha, M., \& Willman, B. 2008 \apj, 678, 614
\bibitem[Strigari et al.(2008b)]{s08b} 
Strigari, L. E., Bullock, J. S., Kaplinghat, M., Simon, J. D, Geha,
M., Willman, B., \& Walker, M. G. 2008, \nat, 454, 1096
\bibitem[Takei et al.(2007)]{tak07}
Takei, Y., et al. 2007, \pasj, 59, S339
\bibitem[Tawara et al.(2008)]{taw08}
Tawara, Y., et al. 2008, \pasj, 60, S307
\bibitem[Tawa et al.(2008)]{ta08}
Tawa, N., et al. 2008, \pasj, 60, S11
\bibitem[Tokoi et al.(2008)]{tok08}
Tokoi, K., et al. 2008, \pasj, 60, S317
\bibitem[Uchiyama et al.(2008)]{u08a}
Uchiyama, Y., et al. 2008, \pasj, 60, S35
\bibitem[Uchiyama et al.(2009)]{u08b}
Uchiyama, Y., et al. 2009, \pasj,, 61, S9
\bibitem[Viel et al.(2008)]{v08}
Viel, M., Becker, G. D., Bolton, J. S., Haehnelt, M. G., Rauch, M., \&
Sargent, W. L. W. 2008, \prl, 100, 041304
\bibitem[Watson et al.(2006)]{w06}
Watson, C. R., Beacom, J. F., Y\"uksel, H., \& Walker, T. P. 2006, \prd,
74, 033009
\bibitem[Werner et al.(2008)]{w08} 
Werner, N., Finoguenov, A., Kaastra, J. S., Simionescu, A., Dietrich,
J. P., Vink, J., \& Bohringer, H. 2008, \aap, 482, 29
\bibitem[Wilkinson et al.(2004)]{w04} 
Wilkinson, M. I., Kleyna, J. T., Evans, N. W., Gilmore, G. F., Irwin,
M. J., \& Grebel, E. K. 2004, \apj, 611, L21
\bibitem[Wu(2007)]{w07} 
Wu, X. 2007, \apj, submitted (astrop-ph/0702233v1)
\bibitem[Yang, Mo, \& Kauffmann(2003)]{ymk03} 
Yang, X. H., Mo, H. J., \& Kauffmann, G. 2003, \mnras, 339, 387
\bibitem[Yaqoob(1998)]{yaq98} 
Yaqoob, T. 1998, \apj, 500, 89
\bibitem[Zaritsky, Gonzalez, \& Zabludoff(2006)]{zgz} 
Zaritsky, D., Gonzalez, A. H., \& Zabludoff, A. 2006, \apj, 638, 725
\end{thebibliography}
\end{document}